\documentclass[runningheads]{llncs}

\usepackage[T1]{fontenc}
\usepackage{graphicx}
\usepackage{booktabs}
\usepackage{amsmath}
\usepackage{amssymb}
\usepackage{tikz}
\usetikzlibrary{arrows.meta,positioning}
\definecolor{cBlue}{HTML}{2A78D6}
\definecolor{cOrange}{HTML}{EB6834}
\definecolor{cAqua}{HTML}{1BAF7A}
\definecolor{cInk}{HTML}{0B0B0B}
\usepackage{xurl}
\usepackage[hidelinks]{hyperref}
\begin{document}

\title{Coding Agents Have Converged: Why the SWE-bench Leaderboard Can No Longer
Order Its Top Entries, and What to Measure Instead}

\titlerunning{Coding Agents Have Converged}
\author{Fengshuo LIU\inst{1}\orcidID{0009-0008-2275-6921}\thanks{Corresponding author.} \and
Ying LIU\inst{2} \and
Ruize SUN\inst{3} \and
Lie LUO\inst{4} \and
Siyuan GUO\inst{4}}
\authorrunning{F. LIU et al.}
\institute{Imperial College London, London, United Kingdom\\
\email{fengshuo.liu26@imperial.ac.uk} \and
The Hong Kong Polytechnic University, Hong Kong SAR\\
\email{yarden.liu@connect.polyu.hk} \and
Korea University, Seoul, Korea\\
\email{2025150487@korea.ac.kr} \and
Jinan University, Guangzhou, China\\
\email{lxl243@stu2022.jnu.edu.cn, kayaking@stu2025.jnu.edu.cn}}

\maketitle

\begin{abstract}
Small differences on coding-agent leaderboards are often read as an ordering
of systems. We audit whether the published verdicts support this reading,
using $254$ SWE-bench submissions across four splits without running models.
On Verified, the leading two entries each resolve $396$ of $500$ instances.
The top ten share $285$ successes and $51$ failures, leaving $164$ instances
that distinguish their outcomes. Frontier solution sets have median nesting
$0.935$ against a score-implied baseline of $0.774$, indicating strongly shared
successes. Scores also depend on the evaluated model--scaffold pair:
observed within-model scaffold ranges reach $29.8$ percentage points, compared
with the $8.8$-point spread of the top thirty. Six of nine cell-mean interaction
tests remain significant after Holm correction, although this observational
design does not identify causal scaffold effects. Exact paired McNemar tests
separate none of the $29$ adjacent Verified top-thirty pairs at
$\alpha=0.05$, while the larger Test split separates $14$ of $23$.
A stated leader-based rule yields three descriptive tiers, or two after Holm
correction; non-rejection does not establish equivalence. We release the
partition and a five-step audit protocol that profiles shared outcomes,
tests paired differences, reports grouping sensitivity, and estimates the
instance budget needed for resolution. The results motivate reporting
comparison-set-specific resolution and model--scaffold provenance instead of
interpreting small aggregate gaps as established rank differences.
\keywords{SWE-bench \and Agent evaluation \and Trustworthy evaluation
\and Responsible data intelligence \and Benchmark saturation}
\end{abstract}

% ===========================================================================
\section{Introduction}
% ===========================================================================

SWE-bench Verified \cite{jimenez2024swebench,openai2024verified} evaluates coding
agents on $500$ real GitHub issues. Each agent produces a patch, which the
official harness applies before running repository tests. The two leading
entries each resolve $396$ instances, and six more sit within $14$ of them.
The leaderboard displays ranks $1$ through $8$, often read as an ordering for
model selection, procurement, and research reporting. Shared instance-level
outcomes let us test whether the evidence supports that reading.

That reading is already contested, but along a different axis. One line of work
asks whether the \emph{tasks} are sound --- automated auditing finds quality
problems in over a quarter of tasks and shows that filtering them shifts rankings
and raises mean Verified scores by $9.9\%$ \cite{wang2026aba}, with related
empirical findings on weak tests and leakage \cite{aleithan2024swebenchplus},
outcome validity \cite{zhu2025bestpractices} and lucky passes
\cite{agentlens2026}, and with arguments that coding benchmarks are misaligned
with agentic software engineering \cite{position2026misaligned}. We ask the other
question: even where the tasks are sound, can the \emph{ranks} be read? This is an
audit of a public evaluation artefact --- its provenance, its curation, and the
trustworthiness of the comparisons it is used to license --- and we measure where
that line falls, so that the designers of the next benchmark, and of the internal
suites now being built to choose models, need not rediscover it.

\begin{figure}[tb]
\centering
\includegraphics[width=\textwidth]{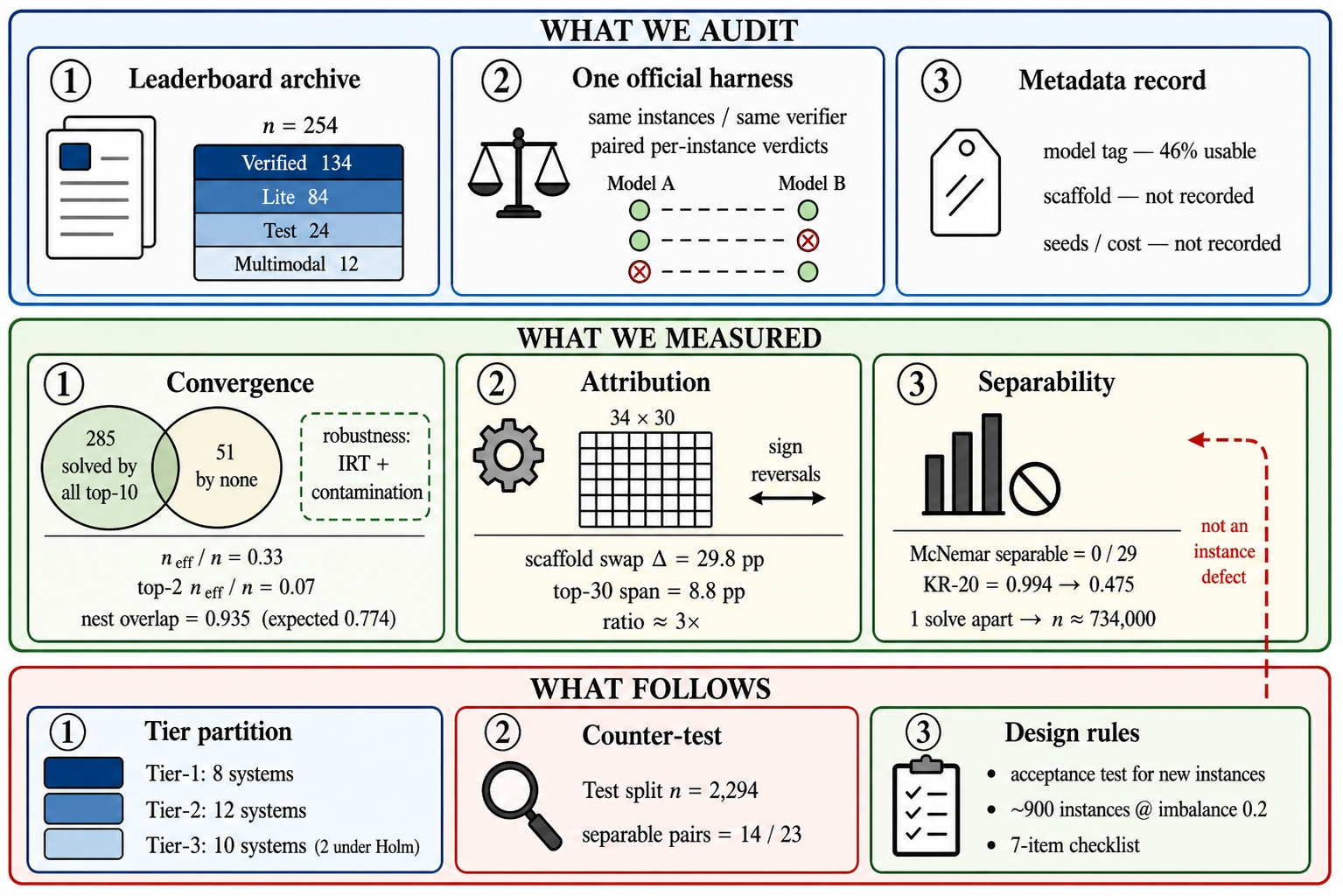}
\caption{The audit: $254$ public submissions and incomplete metadata feed
analyses of shared outcomes, model--scaffold associations, and paired rank
separability. Outputs are descriptive tiers, a Test-split counter-test, and
benchmark-design guidance. Tiers use uncorrected leader comparisons; Holm
gives $19/11$ groups. Attribution is observational, not causal. The $734{,}000$
instance figure is an independent-sample power reference, not a paired-test
requirement (Section~\ref{sec:tiers}).}
\label{fig:overview}
\end{figure}

We find shared frontier successes, model--scaffold associations, and insufficient
paired evidence for adjacent rank differences. The resulting descriptive groups
depend on the stated rule and correction. We then examine implications for
SWE-bench, broader suites, and internal model selection. Figure~\ref{fig:overview}
summarises the argument.

\paragraph{What this paper contributes.}
An audit protocol that mines the per-instance verdict matrix every public
leaderboard already publishes, and turns it into actionable structure
(Protocol~1, Section~\ref{sec:tiers}): a degeneracy profile that localises
where resolution is lost, a nesting coefficient measured against its
score-implied null that identifies convergence as the mechanism, paired
separability testing that yields a publishable tier partition, and an
inversion that prices new instances. Two of its constructs are, to our
knowledge, new: the comparison-set-relative effective size
$n_{\mathit{eff}}(S)$ and the nesting coefficient against its baseline.
Applied to SWE-bench, the protocol returns a far more extreme failure state
than prior audits of other leaderboards, which found unresolvable pairs to be
the exception ($11$ of $40$ and $4$ of $9$ in \cite{kotawala2026resolution});
on the Verified frontier they are $29$ of $29$. The model$\times$scaffold
design of Section~\ref{sec:pair} is reconstructed from public metadata, the
protocol checks its own candidate remedies before recommending them
(Section~\ref{sec:implications}), and we release the full pipeline as a
reusable tool. The statistical machinery itself is established, and we use it
as such: paired adjacent-rank testing and required-sample-size inversion
follow \cite{kotawala2026resolution}, rank intervals
\cite{neuhof2026rankintervals}, benchmark power analysis \cite{card2020power},
reduced-size construction
\cite{polo2024tinybenchmarks,perlitz2024efficient}, item-level psychometrics
\cite{lostinbench2025}, and the information-retrieval work on test collection
reliability \cite{sanderson2005effort,sakai2016topicset,roitero2020fewertopics}
long predates all of it. The task-quality work cited above is complementary
to ours --- it asks whether the tasks are sound; we ask whether the ranks are
--- and we draw on it in Section~\ref{sec:implications}.

% ===========================================================================
\section{Data}\label{sec:data}
% ===========================================================================

For every leaderboard submission the SWE-bench maintainers publish the official
harness verdict on each instance, plus a \texttt{metadata.yaml} carrying a model
tag \cite{swebench2026experiments}; every figure below is computed from the
leaderboard as retrieved on 30 July 2026, with the analysis manifest frozen in
the supplementary package. Table~\ref{tab:pool} summarises the pool: all submissions on the four
public splits that publish per-instance results. Verified and Lite share only
$93$ instances ($31\%$ of Lite), so Lite is a largely independent replication.
The ten excluded Multimodal submissions record \texttt{resolved} as an integer
count rather than a list of instance identifiers, and so cannot support any
per-instance analysis --- a first sign of the metadata problem described next.
Submissions span 2023--2025 and range from $2$ to $396$ resolved instances on
Verified. ``Top $K$'' always means the $K$ highest-scoring submissions; a
missing verdict counts as unresolved, which is the convention the displayed
score already implies. What makes this pool analysable is that every entry was
graded by the same official harness on the same instances, which is what
licenses the paired tests below; every number in this paper can be recomputed
by anyone from the published verdicts.

\begin{table}[tb]
\caption{The analysis pool: the four public splits that publish per-instance
verdicts, as retrieved on 30 July 2026. Sections~\ref{sec:pair}--\ref{sec:tiers}
use Verified; Section~\ref{sec:converge} reports all four.}
\label{tab:pool}
\centering
\small
\setlength{\tabcolsep}{9pt}% wider gutters: column gaps must exceed word spaces
\begin{tabular*}{\textwidth}{@{\extracolsep{\fill}}lrrll@{}}
\toprule
Split & Submissions & Instances & Verified overlap & Role \\
\midrule
Verified   & 134        & 500     & ---      & main analysis \\
Lite       & 84         & 299     & 93 (31\%) & indep.\ replication \\
Test       & 24         & 2{,}294 & ---      & counter-test \\
Multimodal & 12 of 22   & 301     & ---      & descriptive \\
\bottomrule
\end{tabular*}
\end{table}

\paragraph{The provenance record barely supports analysis.}
Only $61$ of $134$ Verified submissions ($46\%$) carry a single usable model tag,
and the tags are inconsistent: \texttt{claude-\allowbreak sonnet-\allowbreak 4}
and \texttt{claude-\allowbreak 4-\allowbreak sonnet} denote the same model, date
suffixes appear and
disappear, and some entries give a URL instead of an identifier. We normalised
identifiers by hand and derived the scaffold from the submission name, yielding
$34$ scaffolds, $30$ models and $55$ occupied cells, five of which hold more than
one submission. A leaderboard whose purpose is comparison does not record what
was compared in machine-readable form; Section~\ref{sec:implications} returns to
this.

For even samples, descriptive score-range, nesting and budget medians use the
upper middle observation; the IRT analysis uses midpoint medians.

% ===========================================================================
\section{The Systems Solve the Same Problems}\label{sec:converge}
% ===========================================================================

\subsection{Most of the Benchmark Cannot Separate the Leaders}

For a comparison set $S$ of systems, call an instance \emph{degenerate} if every
member of $S$ resolves it or none does: it cannot contribute to any comparison
within $S$, yet it contributes $1/n$ to every score. The effective size of a
comparison is the number of non-degenerate instances:
\begin{equation}
n_{\mathit{eff}}(S) \;=\; \bigl|\{\, i : 0 < \text{resolved}_S(i) < |S| \,\}\bigr|
\end{equation}
Following the test-collection tradition \cite{sakai2016topicset,roitero2020fewertopics}, this
is the sample size that matters, and it is a property of the pair (benchmark,
comparison set) rather than of the benchmark alone.

\begin{figure}[tb]
\centering
\includegraphics[width=\textwidth]{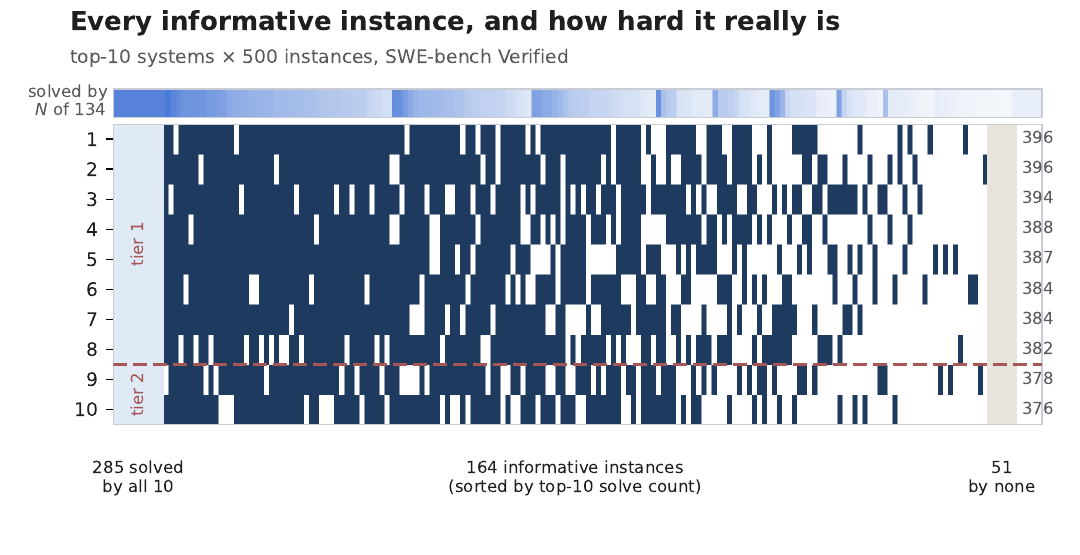}
\caption{The $164$ instances that still separate the top ten, one row per
system, with the two degenerate blocks compressed to side strips ($285$
resolved by all ten, $51$ by none). The top strip counts how many of
\emph{all} $134$ submissions resolve each instance --- a difficulty gradient
the top ten themselves cannot see. The dashed line marks the tier-1/tier-2
boundary (Section~\ref{sec:tiers}).}
\label{fig:matrix}
\end{figure}

Table~\ref{tab:neff} reports it. Over all $134$ Verified submissions
$n_{\mathit{eff}}/n = 0.94$ and the benchmark looks healthy. Over the top ten it
is $0.33$: $285$ instances are solved by all ten, $51$ by none, and $164$ remain.
Over the top two it is $0.07$ --- $36$ instances. Figure~\ref{fig:matrix} shows
the same decomposition instance by instance. Lite shows the same pattern
from a less saturated base ($0.89 \to 0.54$ at the top ten); that its frontier
value is higher is what one expects of a harder benchmark, so the quantity tracks
saturation sensibly. The collapse holds on all four splits
that publish per-instance results (Table~\ref{tab:splits}), including the full
$2294$-instance Test set.

\begin{table}[tb]
\caption{Degenerate and discriminating instances as the comparison set narrows.
KR-20 is omitted for $K<10$, where it is unstable.}
\label{tab:neff}
\centering
\small
\setlength{\tabcolsep}{9pt}% wider gutters: column gaps must exceed word spaces
\begin{tabular*}{\textwidth}{@{\extracolsep{\fill}}lrrrr@{}}
\toprule
$K$ & solved by all & solved by none & $n_{\mathit{eff}}/n$ & KR-20 \\
\midrule
\multicolumn{5}{@{}l}{\emph{Verified} ($134$ submissions, $n=500$; top-10 mean $77.3\%$)}\\
2   & 378 & 86 & 0.07 & --- \\
5   & 334 & 60 & 0.21 & --- \\
10  & 285 & 51 & 0.33 & 0.475 \\
20  & 250 & 41 & 0.42 & 0.722 \\
50  & 146 & 35 & 0.64 & 0.940 \\
134 & \multicolumn{3}{c}{$n_{\mathit{eff}}/n = 0.94$} & 0.994 \\
\midrule
\multicolumn{5}{@{}l}{\emph{Lite} ($84$ submissions, $n=299$; top-10 mean $54.3\%$)}\\
10  & \phantom{0}73 & 65 & 0.54 & 0.867 \\
20  & \phantom{0}35 & 49 & 0.72 & 0.874 \\
84  & \multicolumn{3}{c}{$n_{\mathit{eff}}/n = 0.89$} & 0.982 \\
\bottomrule
\end{tabular*}
\end{table}

\subsection{Solutions Are Nested, Not Complementary}\label{sec:nesting}

Degeneracy could in principle coexist with specialisation: systems might solve
overlapping-but-different sets. They do not. For a pair with $|A| \ge |B|$ define
the nesting coefficient and its score-implied baseline:
\begin{equation}
\mathrm{cov}(A,B) \;=\; \frac{|A \cap B|}{|B|},
\qquad
\mathbb{E}[\,\mathrm{cov}\,] \;=\; \frac{|A|}{n}
\end{equation}
--- the fraction of the weaker system's solutions that the stronger one also
produces ($\mathrm{cov}=1$ means $B \subseteq A$); the baseline follows from
allocating instances at random subject to the observed scores, and gives a
reference for each ability band.

\begin{table}[tb]
\caption{Nesting of solution sets within ability bands (SWE-bench Verified),
against the baseline implied by the scores alone.}
\label{tab:nesting}
\centering
\small
\setlength{\tabcolsep}{9pt}% wider gutters: column gaps must exceed word spaces
\begin{tabular*}{\textwidth}{@{\extracolsep{\fill}}lrrrr@{}}
\toprule
band (resolved instances) & systems & median $\mathrm{cov}$ & baseline & excess \\
\midrule
frontier ($\ge 370$)      & 16 & 0.935 & 0.774 & $+0.161$ \\
strong ($300$--$369$)     & 38 & 0.913 & 0.708 & $+0.205$ \\
mid ($200$--$299$)        & 48 & 0.828 & 0.530 & $+0.298$ \\
weak ($100$--$199$)       & 23 & 0.712 & 0.370 & $+0.342$ \\
\bottomrule
\end{tabular*}
\end{table}

Table~\ref{tab:nesting} shows median nesting of $0.935$ at the frontier: when two
leading systems differ, the weaker one's successes are almost entirely a subset
of the stronger one's. Nesting exceeds its score-implied baseline in every band,
so this is not an artefact of high scores; and while the \emph{excess} over
baseline is largest in the weakest band ($+0.342$ against $+0.161$ at the
frontier), absolute nesting peaks exactly where the ranking is read --- the
frontier --- which is what removes resolution there.

The practical consequence is that there is little complementarity to exploit. The
union of instances solved by the top two is $414$ against $396$ for the best
single system, and by the top ten $449$ --- gains, but small ones, and obtained
by combining systems rather than by ordering them.

\subsection{Item Parameters, Data Quality, and Contamination}\label{sec:robust3}

Three alternative readings of this section can be tested rather than merely
acknowledged.

\paragraph{An item-response view, and why $n_{\mathit{eff}}$ is still needed.}
A two-parameter logistic model fitted to the $134\times468$ non-constant response
matrix converges cleanly (in-sample accuracy $0.868$ against a base rate of
$0.550$; $\mathrm{corr}(\theta,\text{score})=0.971$; discrimination bounded to a
conventional $[0.1,4]$). Its test information function peaks at $\theta=-0.62$,
\emph{below} the median system: the instrument carries $2.3\times$ more
information about a median system ($I=278$, ability $\mathrm{SE}=0.060$) than
about one in the top-ten band ($I=118$, $\mathrm{SE}=0.092$). Item analysis alone does not expose the problem: instances degenerate for the
top ten have \emph{higher} mean discrimination than the informative ones ($2.52$
vs $1.82$), because they separate strongly across the full 2023--2025 range while
telling one 2025 system from another not at all. Discrimination is a property of
an item against a population; $n_{\mathit{eff}}$ is defined against a chosen
comparison set, and it is the comparison set that has moved.

\paragraph{Degeneracy is not simply task defects.}
Audits find quality problems in a substantial share of agentic benchmark
instances \cite{wang2026aba}, raising the possibility that degenerate instances are broken rather than shared.
The human effort estimates released with SWE-bench Verified argue otherwise: of
the $285$ instances every top-ten system resolves, $49\%$ were rated ``$<15$
min'' and none ``$>4$ hours''; of the $51$ none resolves, $14\%$ were rated
``$<15$ min'' and $33\%$ over an hour; the informative middle sits between them
at every level. That ordering comes from annotators working before any of these
systems existed, and is not what a defect account predicts.

\paragraph{No detectable age effect.}
If contamination drove degeneracy it should favour older pull requests, but across
the $500$ instances the correlation between PR year and top-ten solve rate is
$-0.069$ (95\% bootstrap CI $[-0.148,+0.012]$): no age effect is detectable.

% ===========================================================================
\section{The Number Is Not a Property of the Model}\label{sec:pair}
% ===========================================================================

A leaderboard entry is produced by a \emph{pair}: a model, and the scaffold ---
the agent loop, tool set, retrieval and control policy --- that drives it. The
leaderboard displays neither factor as such. Reconstructing the pair from the
public metadata (Section~\ref{sec:data}) lets us ask how much each contributes.

\subsection{Holding the Model Fixed}

Among models appearing with at least two scaffolds, the observed within-model
scaffold range has a median of $78$ instances ($15.6$pp). For
\texttt{claude-3-5-sonnet}, which appears with nine scaffolds, the range is $168$
to $317$ --- $149$ instances, or $29.8$pp. For \texttt{claude-4-sonnet} (eight
scaffolds) it is $99$ instances; for \texttt{gpt-4o} (six) it is $78$.
Symmetrically, the observed within-scaffold model range has a median of $60$ instances ($12.0$pp). For scale, \textbf{the entire spread of the
top thirty submissions is $8.8$pp}.

Five cells contain repeat submissions of the same scaffold with the same model at
different dates; their ranges are $13$, $15$, $27$, $41$ and $116$ instances
(median $27$, i.e.\ $5.4$pp) --- a replication floor that the median within-model
scaffold spread exceeds by $2.9\times$. The largest case, three
\texttt{epam-ai-run} $\times$ \texttt{claude-3-5-sonnet} submissions scoring
$198$, $277$ and $314$, shows that a scaffold is itself a moving target between
submission dates. We average replicates throughout.

A least-squares additive fit $y = \mu + \alpha_{\text{scaffold}} +
\beta_{\text{model}}$ on the connected core ($20$ cells, $10$ scaffolds, $7$
models) gives $R^2 = 0.989$ with a fitted scaffold-effect range of $37.6$pp
against $47.9$pp for models, a ratio of $0.78$. The reading is not that scaffolds
matter more than models, but that the fitted ranges are of the same order.
These are observational associations on a sparse core, not causal effects or
fractions of variance explained; team effort and co-optimisation are confounded.

\subsection{The Two Factors Do Not Separate}

For scaffolds $s_1,s_2$ and models $m_1,m_2$ with all four cells present, write
$\delta(m) = \text{score}(s_1,m) - \text{score}(s_2,m)$; the interaction is
$\delta(m_1) - \delta(m_2)$. All submissions are graded on the same instances, so
we test it as a paired difference-in-differences, bootstrapping over the $500$
instances, averaging submissions within each cell. With $100{,}000$ paired
bootstrap resamples, \textbf{six of nine interactions remain significant after
Holm correction} across all nine tests ($\alpha=0.05$); excluding the 2023
\texttt{rag} case leaves five significant cases among the other eight under
that same correction. We report two-sided bootstrap-tail probabilities with a
finite-resample correction. Using the best submission instead gives six
uncorrected and five Holm-significant cases; the mean is our primary analysis.

\texttt{epam-ai-run} leads \texttt{sweagent} by $95$ instances at
\texttt{claude-3-5-sonnet} and $51$ at \texttt{claude-4-sonnet}
(interaction $+44$, Holm-adjusted $p=0.0035$).
\texttt{agentless} versus \texttt{epam-ai-run} \emph{reverses sign} between
\texttt{claude-3-5-sonnet} and \texttt{gpt-4o}. The interaction is $-75.5$
instances ($-15.1$pp). The \texttt{autocoderover} comparison with
\texttt{epam-ai-run} also reverses ($-83$ instances).
Both reversals have adjusted $p<0.001$. The observed scaffold ordering
thus depends on the model. A fixed scaffold correction cannot recover a model
ranking from these submissions.

% ===========================================================================
\section{Descriptive Tiers, Not a Strict Ranking}\label{sec:tiers}
% ===========================================================================

Because every submission is graded on the same instances, adjacent entries should
be compared with a paired test. Following the paired-resolution analysis of
\cite{kotawala2026resolution}, we apply an exact McNemar test to each adjacent
pair in the top thirty:
\textbf{on Verified none of the $29$ pairs is separable at $\alpha=0.05$}, and
the same holds on Lite. The median gap is one instance and the median discordance
$61$ (Verified) and $57$ (Lite); widening to rank distances of two through five
on Verified leaves the count at zero.

The counter-test on the other three splits is what makes this a diagnosis rather
than a blanket claim. Table~\ref{tab:splits} runs the same test everywhere
per-instance results exist. Where the leading submissions still span a wide
range of ability and the instance count is large
--- the $2294$-instance Test split, whose top $24$ range from $0.2\%$ to
$52.6\%$ --- $14$ of $23$ adjacent pairs \emph{are} separable, at a median gap of
$42$ instances. The same statistic that finds nothing on Verified finds plenty
where ability still spans a range: irresolvability is a property of a converged
comparison set, not of the benchmark family, and the split-by-split pattern is
the empirical counterpart of the design arithmetic in
Section~\ref{sec:implications}.

\begin{table}[tb]
\caption{The same adjacent-pair test on every public split with per-instance
results. $n_{\mathit{eff}}/n$ and ``spread'' are computed over the top ten and
the tested comparison set respectively. Separability tracks the spread and the
instance count, exactly as the design arithmetic of
Section~\ref{sec:implications} predicts.}
\label{tab:splits}
\centering
\small
\setlength{\tabcolsep}{9pt}% wider gutters: column gaps must exceed word spaces
\begin{tabular*}{\textwidth}{@{\extracolsep{\fill}}lrrrrrr@{}}
\toprule
split & $n$ & systems & $n_{\mathit{eff}}/n$ & spread & adj.\ pairs & separable \\
\midrule
Verified   & \phantom{0}500 & 134 & 0.33 & \phantom{0}8.8\,pp & 29 & \phantom{0}0 \\
Lite       & \phantom{0}299 & \phantom{0}84 & 0.54 & 21.1\,pp & 29 & \phantom{0}0 \\
Multimodal & \phantom{0}301 & \phantom{0}12 & 0.44 & 18.3\,pp & 11 & \phantom{0}0 \\
Test       & 2294 & \phantom{0}24 & 0.52 & 52.4\,pp & 23 & 14 \\
\bottomrule
\end{tabular*}
\end{table} As an independent-sample reference, detecting a $0.2$pp gap
(one of $500$ instances) at a $75\%$ baseline and $80\%$ power requires about
$734{,}000$ instances per system. This two-proportion calculation is not the
paired McNemar requirement; Section~\ref{sec:implications} gives paired design
arithmetic without an $80\%$ power guarantee.

\begin{figure}[tb]
\centering
\includegraphics[width=\textwidth]{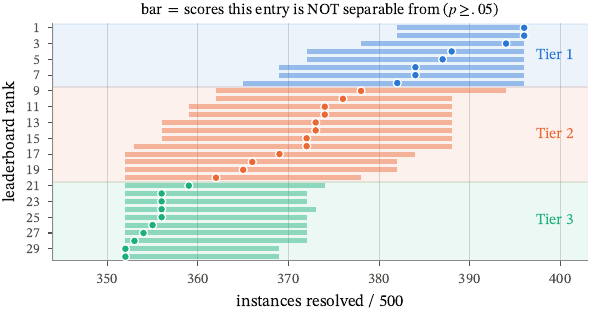}
\caption{The top thirty of SWE-bench Verified. Each dot is a submission's score;
the grey bar spans the scores of every other entry in the top thirty from which
an exact McNemar test cannot separate it ($p \ge 0.05$). The bars are wide enough
to cover much of the table. Bars are score spans, not confidence or rank
intervals; non-significance need not hold for every intermediate score. Shaded
bands mark the three uncorrected leader-based tiers.}
\label{fig:tiers}
\end{figure}

Non-rejection does not establish equivalence or non-inferiority: we specify no
practical equivalence margin. Non-significance is also non-transitive. Our tiers
are descriptive, not simultaneous rank intervals \cite{neuhof2026rankintervals}.
For Table~\ref{tab:tiers}, sort by score (ties by submission identifier), join the
current tier when its leader's uncorrected McNemar $p\ge0.05$, and otherwise
start a new tier. Figure~\ref{fig:tiers} shows this convention, not equal ability.

\begin{table}[tb]
\caption{Descriptive tiers of the SWE-bench Verified top thirty using
uncorrected leader comparisons at $\alpha=0.05$. Membership does not establish
equivalence or that every within-tier pair is non-significant. Boundaries depend
on the rule; full membership is released with the analysis scripts.}
\label{tab:tiers}
\centering
\small
\setlength{\tabcolsep}{9pt}% wider gutters: column gaps must exceed word spaces
\begin{tabular*}{\textwidth}{@{\extracolsep{\fill}}clr@{}}
\toprule
Tier & resolved instances (\% of $500$) & systems \\
\midrule
1 & $382$--$396$ \;($76.4$--$79.2\%$) & 8 \\
2 & $362$--$378$ \;($72.4$--$75.6\%$) & 12 \\
3 & $352$--$359$ \;($70.4$--$71.8\%$) & 10 \\
\bottomrule
\end{tabular*}
\end{table}

\paragraph{Sensitivity to grouping and correction.}
Across the $435$ pairs in
the top thirty, $194$ ($44.6\%$) are separable uncorrected and $41$ ($9.4\%$)
survive Holm--Bonferroni over the whole family --- but among the $29$
\emph{adjacent} pairs the count is $0$ either way, and the smallest adjacent
$p$-value is $0.545$, far from any threshold a correction could matter at.
Requiring non-significance against every current tier member gives the same
$8/12/10$ grouping here. With Holm-adjusted $p$-values, both rules instead give
$19/11$ groups. Thus two or three groups arise under these four specifications;
this is not a universal upper bound on distinguishable capability levels.

Internal consistency agrees. KR-20 over the system$\times$instance matrix is
$0.994$ across all $134$ Verified submissions and $0.940$ across the top fifty,
but $0.722$ across the top twenty and $0.475$ across the top ten
(Table~\ref{tab:neff}). Against the conventional thresholds of roughly $0.90$ for
individual-level decisions and $0.70$ for group-level research use
\cite{nunnally1994psychometric}, the instrument
has much higher internal consistency across the full pool than at the frontier.
These thresholds are descriptive references, not a validation of our tier
boundaries or evidence of equivalence within tiers.

\smallskip
\noindent\fbox{\begin{minipage}{\dimexpr\textwidth-2\fboxsep-2\fboxrule}
\smallskip
\textbf{Protocol 1 (resolution audit).} \emph{Input:} the per-instance verdict
matrix produced by one harness, and the leaderboard order. \emph{Mine it in
five steps.}
\begin{itemize}
\setlength{\itemsep}{1pt}
\item[(1)] \textbf{Profile}: compute the effective size
$n_{\mathit{eff}}(S)$ over nested comparison sets; where the profile
collapses, ranks stop being readable.
\item[(2)] \textbf{Explain}: the nesting
coefficient against its score-implied null to quantify shared successes.
\item[(3)] \textbf{Test}: exact paired
McNemar for every pair; report raw and Holm-adjusted $p$-values over the
declared comparison family.
\item[(4)] \textbf{Partition}:
walk the sorted ranking, comparing each entry to the current tier leader;
start a new tier at $p<0.05$. State whether raw or adjusted $p$-values are used,
and report sensitivity to requiring all current members to be non-significant.
Groups do not establish equivalence.
\item[(5)] \textbf{Budget}: invert the paired
condition to price new instances, checking candidate remedies before
recommending them.
\end{itemize}
\emph{Output:} a tier partition, an acceptance criterion
for new instances, and an instance budget. Every step consumes only published
verdicts; the released pipeline reproduces every number in this paper.
\smallskip
\end{minipage}}

% ===========================================================================
\section{Design Implications}\label{sec:implications}
% ===========================================================================

The three findings point to concrete changes.

\subsection{For SWE-bench and Benchmarks Like It}

\begin{figure}[tb]
\centering
\includegraphics[width=\textwidth]{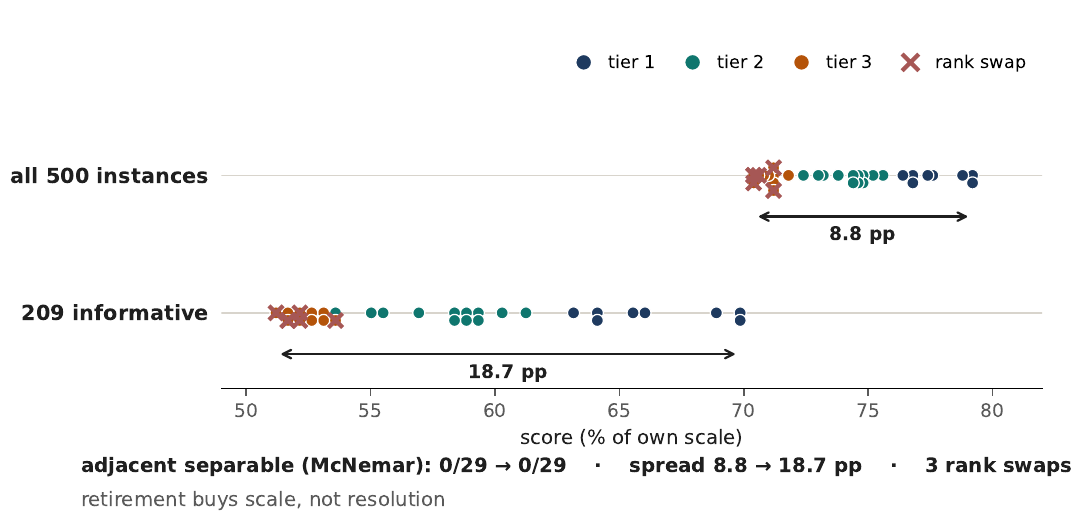}
\caption{Rescoring the top thirty on only the $209$ instances non-degenerate for
the top twenty. The scale widens and three adjacent pairs swap places, yet nothing
is gained: the paired tests behind Section~\ref{sec:tiers}'s tiers already discard
degenerate instances, so separable pairs stay at zero.}
\label{fig:retire}
\end{figure}

\textbf{Report $n_{\mathit{eff}}$ --- but do not expect retirement to buy
resolution.} An instance solved by every system in the frontier set contributes
nothing to ordering it, and $285$ of $500$ are in that state for the top ten
(Section~\ref{sec:converge}). Retiring them is the tempting fix, and it does not
work: Figure~\ref{fig:retire} rescales the top thirty onto the $209$ instances
non-degenerate for the top twenty, and the visible spread more than doubles, from
$8.8$ to $18.7$ percentage points, reordering three adjacent pairs --- yet
separable pairs stay at $0$ of $29$, because a paired test already ignores
instances both systems agree on. Retirement is worth doing for evaluation
\emph{cost} and to stop scores drifting into a compressed range that invites false
precision, but it adds no power, and advertising the wider spread as sharper
discrimination would mislead readers. What buys resolution is instances that break
the nesting of Section~\ref{sec:nesting}, discussed next.

\textbf{Add instances that break the nesting, not instances that add count.}
Because solution sets are nested (Section~\ref{sec:nesting}), more instances of
the kind already present mostly add degenerate ones. Separating a paired
comparison prices directly against the discordant instances
\cite{kotawala2026resolution}:
\begin{equation}
|b-c| \;\gtrsim\; 1.96\sqrt{b+c},
\qquad
k \;\geq\; 1.96^{2}\,\frac{b+c}{(b-c)^{2}}
\end{equation}
--- the gap a benchmark must show, and the scale factor $k$ needed at fixed
discordance and imbalance rates. For the ten leading adjacent pairs on
Verified, excluding two zero-gap pairs, the upper median multiplier is $52\times$ --- roughly $26{,}000$ instances of
the same character. At an unchanged zero imbalance, additional instances do
not separate the two tied pairs. Instances that disagree \emph{and} lean one way
change the picture: for the reference pair with $54/500$ discordance, an
imbalance rate of $0.2$ needs about $900$ instances rather than $26{,}000$. That ratio, not a target task count,
is the curation criterion, and it gives a measurable acceptance test for a
candidate instance. The $26{,}000$ estimate concerns paired significance at
observed rates; the $734{,}000$ reference assumes independent samples and
$80\%$ power (Section~\ref{sec:tiers}). They are not directly comparable.

\textbf{Record the pair in machine-readable form.} Since scaffold and model
contribute comparably (Section~\ref{sec:pair}) and $54\%$ of Verified submissions
cannot currently be placed in a factorial at all, a structured
$(\text{model},\text{scaffold},\text{version})$ field --- plus attempts, and
whether the harness was modified --- would let anyone reproduce
Section~\ref{sec:pair} without hand normalisation. Provenance of this kind is the
cheapest governance change proposed here and the prerequisite for the rest.

\textbf{Publish tiers with the linkage rule, not strict ranks.} A ranked list
asserts an order the data does not contain. Tiers, or the rank intervals of
\cite{neuhof2026rankintervals}, state what is supported.

\textbf{Breadth does not by itself restore resolution.} WorkBuddy Bench has
code ($80$ tasks), web ($70$), security ($60$) and office ($50$)
subsets~\cite{workbuddybench}. Under independent-sample reasoning at $80\%$
power and a $65\%$ baseline, detectable differences are roughly $23$pp at
$n=50$, $19$pp at $n=80$, and $8.2$pp at $n=500$. Pairing changes these
requirements. \emph{Judgement:} report per-subset resolution and grouping
sensitivity, testing whether pooled results support any finer ranking claims.

\subsection{For Organisations Building an Internal Benchmark to Choose a Model}

\begin{enumerate}
\item \textbf{Evaluate the deployed pair.} Observed scaffold orderings reverse
across models (Section~\ref{sec:pair}); evaluate $(M,\text{your harness})$
rather than transferring an external model ranking.

\item \textbf{Size the benchmark for the decision.} At $n=100$, the independent-
sample detectable difference is roughly $17$pp. Paired designs use observed
discordance to refine the budget \cite{card2020power,kotawala2026resolution}.
State the detectable gap before testing; small suites need not resolve $2$pp.

\item \textbf{Track candidate-specific $n_{\mathit{eff}}$.} Tasks all candidates
pass or fail do not distinguish them. Monitor the informative fraction
(Table~\ref{tab:neff}) when curating new tasks.

\item \textbf{Grade from machine-readable artefacts, never from the agent's own
report.} Audits find defects in a substantial share of agentic benchmark items
\cite{wang2026aba,zhu2025bestpractices} and document lucky passes
\cite{agentlens2026}; a harness that reads a verdict the agent prints rather than
the grader's own output detects neither. \emph{Judgement:} persist the patch and
the grader's structured report per task, reject patches touching test files, and
treat any run not recomputable from stored artefacts as missing, not failed.

\item \textbf{Fix the scoring convention for missing verdicts before you run, and
report both.} Whether a timed-out run counts as a failure or is excluded moves a
score materially, and the choice is invisible in a single number.
\emph{Judgement:} pre-register it and publish the count of missing verdicts
beside the score.

\item \textbf{Report tiers to decision-makers.} Given item~2, the honest output
is usually a few tiers plus cost and latency per tier, not a ranked table.
\emph{Judgement:} a rank invites a decision the measurement cannot support.

\item \textbf{Prefer internal tasks, and re-mine them.} Public benchmark
instances are exposed to pretraining overlap \cite{liang2025illusion}; tasks
mined from private repositories start largely free of it, a genuine advantage of
an internal suite. \emph{Judgement:} re-mine periodically, because
an internal suite saturates for the same reason a public one does.
\end{enumerate}

% ===========================================================================
\section{Limitations}\label{sec:limits}
% ===========================================================================

\textbf{The factorial design is observational.} Teams choose scaffold and model
together, so scaffold effects absorb co-optimisation, engineering effort and
differential model access; they are not causal, and $54\%$ of submissions cannot
be placed in the design, which also skews it towards model generations that
several teams have built around. The supported conclusion is the one we draw ---
the displayed number is not attributable to the model alone --- and the effect
remains large where the design is most current: $99$ instances between scaffolds
for \texttt{claude-\allowbreak 4-\allowbreak sonnet}, over twice the top-thirty
spread. A controlled factorial running the same scaffolds over the same models
under one harness would be decisive and is future work.

\textbf{One run per submission; contamination.} Between-system differences
cannot be separated from run-to-run variance. Our intervals and tests condition
on the observed submissions and resample instances; they omit execution
variability and do not quantify total uncertainty or prove equivalence. Verified also overlaps pretraining data: given only the issue
text, models identify the buggy file at $76\%$ accuracy on Verified but $53\%$
on unseen repositories \cite{liang2025illusion}. Our age-based test
(Section~\ref{sec:robust3}) detects no effect, but age is only a proxy, so we do
not read the cohort movement of $n_{\mathit{eff}}/n$ ($0.06 \to 0.50 \to 0.33$
across the 2023--2025 within-cohort top tens) as evidence about capability
growth.

\textbf{Task defects and scope.} We take harness verdicts as given. The
alignment with human effort estimates (Section~\ref{sec:robust3}) argues against
defects being the main driver of degeneracy; where defects do exist
\cite{wang2026aba}, the affected instances still occupy every score's
denominator, so the measurement stands. The pool is one benchmark family, one
harness, and voluntarily submitted. The protocol can consume other paired
verdict matrices, but validation on unrelated leaderboards remains future work;
we do not claim that they share the measured magnitudes.

% ===========================================================================
\section{Conclusion}
% ===========================================================================

Coding agents at the top of SWE-bench Verified have converged: they solve the
same $285$ of $500$ instances, fail the same $51$, and their solution sets are
nested at $0.935$ against a score-implied $0.774$. The number that separates them
is not a property of the model alone: observed within-model scaffold ranges
reach $29.8$pp, more than the top-thirty spread, with model-dependent orderings.
Our rule gives three descriptive tiers, or two after Holm correction; neither
is evidence of within-tier equivalence or a unique capability partition.

SWE-bench detects many differences across a broader capability range, and its practice of
publishing per-instance verdicts for every submission --- not universal among
leaderboards --- is exactly what made this audit possible. It is being read at a
resolution it does not have. The fixes are within reach: report
$n_{\mathit{eff}}$, record the model--scaffold pair, publish tiers, and accept
new instances by what they add to the discordant budget. Applying the protocol
to multi-domain suites and internal benchmarks is a concrete next step; its
empirical validation here is limited to the SWE-bench family.

\paragraph{Reproducibility.} All inputs are the per-instance evaluation results
and metadata that the SWE-bench maintainers publish for every leaderboard
submission; no model access, API keys or private data are involved. The accompanying reproducibility package contains the frozen inputs, analysis
scripts, normalised factorial design, tier membership, and the camera-ready
interaction audit with raw and adjusted $p$-values. Seeds and input hashes
are recorded. Code and reproducibility materials:
\url{https://github.com/Adkid-Zephyr/resolution-audit}.

\bibliographystyle{splncs04}
\bibliography{refs}

\begin{thebibliography}{10}
\providecommand{\url}[1]{\texttt{#1}}
\providecommand{\urlprefix}{URL }
\providecommand{\doi}[1]{https://doi.org/#1}

\bibitem{aleithan2024swebenchplus}
Aleithan, R., Xue, H., Mohajer, M.M., et~al.: {SWE-Bench+}: Enhanced coding
  benchmark for {LLMs}. arXiv preprint arXiv:2410.06992  (2024)

\bibitem{card2020power}
Card, D., Henderson, P., Khandelwal, U., et~al.: With little power comes great
  responsibility. In: Proc. EMNLP. pp. 9263--9274 (2020).
  \doi{10.18653/v1/2020.emnlp-main.745}

\bibitem{position2026misaligned}
Gorinova, M.I., Baker, M., Heineike, A., et~al.: Position: Coding benchmarks
  are misaligned with agentic software engineering. arXiv preprint
  arXiv:2606.17799  (2026)

\bibitem{jimenez2024swebench}
Jimenez, C.E., Yang, J., Wettig, A., et~al.: {SWE-bench}: Can language models
  resolve real-world {GitHub} issues? In: International Conference on Learning
  Representations (ICLR) (2024)

\bibitem{kotawala2026resolution}
Kotawala, A.: Resolution diagnostics for paired {LLM} evaluation. arXiv
  preprint arXiv:2605.30315  (2026)

\bibitem{liang2025illusion}
Liang, S., Garg, S., Zilouchian~Moghaddam, R.: The {SWE-Bench} illusion: When
  state-of-the-art {LLMs} remember instead of reason. In: Proc. ICSE-SEIP. pp.
  395--405 (2026). \doi{10.1145/3786583.3786882}

\bibitem{neuhof2026rankintervals}
Neuhof, B., Benjamini, Y.: Rank intervals for leaderboards: A hierarchical
  framework for model evaluation. arXiv preprint arXiv:2606.08679  (2026)

\bibitem{nunnally1994psychometric}
Nunnally, J.C., Bernstein, I.H.: Psychometric Theory. McGraw-Hill, 3 edn.
  (1994)

\bibitem{openai2024verified}
{OpenAI}: Introducing {SWE-bench} verified.
  \url{https://openai.com/index/introducing-swe-bench-verified/} (2024),
  accessed 30 July 2026

\bibitem{perlitz2024efficient}
Perlitz, Y., Bandel, E., Gera, A., et~al.: Efficient benchmarking (of language
  models). In: Proc. NAACL. pp. 2519--2536 (2024).
  \doi{10.18653/v1/2024.naacl-long.139}

\bibitem{polo2024tinybenchmarks}
Polo, F.M., Weber, L., Choshen, L., et~al.: {tinyBenchmarks}: Evaluating {LLMs}
  with fewer examples. In: Proc. ICML. pp. 34303--34326. PMLR 235 (2024)

\bibitem{roitero2020fewertopics}
Roitero, K., Culpepper, J.S., Sanderson, M., et~al.: Fewer topics? a million
  topics? both?! on topics subsets in test collections. Inf. Retr. J.
  \textbf{23}(1),  49--85 (2020). \doi{10.1007/s10791-019-09357-w}

\bibitem{agentlens2026}
Sahoo, P., Mittal, G., Li, X., et~al.: {AgentLens}: Revealing the lucky pass
  problem in {SWE-Agent} evaluation. arXiv preprint arXiv:2605.12925  (2026)

\bibitem{sakai2016topicset}
Sakai, T.: Topic set size design. Inf. Retr. J.  \textbf{19}(3),  256--283
  (2016). \doi{10.1007/s10791-015-9273-z}

\bibitem{sanderson2005effort}
Sanderson, M., Zobel, J.: Information retrieval system evaluation: Effort,
  sensitivity, and reliability. In: Proc. ACM SIGIR. pp. 162--169 (2005).
  \doi{10.1145/1076034.1076064}

\bibitem{swebench2026experiments}
{SWE-bench Team}: {SWE-bench} experiments: Open-sourced predictions, execution
  logs, trajectories, and evaluation results.
  \url{https://github.com/swe-bench/experiments} (2026), snapshot retrieved 30
  July 2026; the analysis manifest is frozen in the supplementary artefact

\bibitem{workbuddybench}
{Tencent WorkBuddy Bench Team}: {WorkBuddy Bench}: A multi-domain coding-agent
  benchmark with contamination-resistant task construction. arXiv preprint
  arXiv:2607.20911  (2026), dataset:
  \url{https://huggingface.co/datasets/tencent/workbuddy-bench}

\bibitem{wang2026aba}
Wang, J., Bianchi, F., Zhu, S., et~al.: Automated benchmark auditing for {AI}
  agents and large language models. arXiv preprint arXiv:2605.26079  (2026)

\bibitem{lostinbench2025}
Zhou, H., Huang, H., Zhao, Z., et~al.: Lost in benchmarks? rethinking large
  language model benchmarking with item response theory. arXiv preprint
  arXiv:2505.15055  (2025)

\bibitem{zhu2025bestpractices}
Zhu, Y., Jin, T., Pruksachatkun, Y., et~al.: Establishing best practices for
  building rigorous agentic benchmarks. arXiv preprint arXiv:2507.02825  (2025)

\end{thebibliography}

\end{document}